\documentclass[final,3p,times,twocolumn,authoryear]{elsarticle}
\usepackage{amssymb}
\usepackage{amsmath}
\usepackage{latexsym}
\usepackage{color}
\usepackage{hyperref}
\journal{Advances in Space Research}

\begin{document}

\begin{frontmatter}


%% \author{I.~Zhelyazkov\corref{cor1}\fnref{label2}}

\title{Kelvin--Helmholtz instability in solar H$\alpha$ surges}


\author{I.~Zhelyazkov$\,^{\rm a}$\corref{cor1}}
\address{$^{\rm a}\,$Faculty of Physics, Sofia University, 1164 Sofia, Bulgaria}
\cortext[cor1]{Corresponding author}
\ead{izh@phys.uni-sofia.bg}
\author{T.~V.~Zaqarashvili$\,^{\rm b,c}$, R.~Chandra$\,^{\rm d}$, A.~K.~Srivastava$\,^{\rm e}$, T. Mishonov$\,^{\rm a}$}

\address{$^{\rm b}\,$Space Research Institute, Austrian Academy of Sciences, 8042 Graz, Austria}
\address{$^{\rm c}\,$Abastumani Astrophysical Observatory at Ilia State University, 0162 Tbilisi, Georgia}
\address{$^{\rm d}\,$Department of Physics, DSB Campus, Kumaun University, Nainital 263\,002, India}
\address{$^{\rm e}\,$Department of Physics, Indian Institute of Technology (Banaras Hindu University), Varanasi 221\,005, India}

\begin{abstract}
We study the evolutionary conditions for Kelvin--Helmholtz (KH) instability in a H$\alpha$ solar surge observed in NOAA AR 8227 on 1998 May 30.  The jet with speeds in the range of
$45$--$50$~km\,s$^{-1}$, width of $7$~Mm, and electron number density of $3.83\times 10^{10}$~cm$^{-3}$ is assumed to be confined in a twisted magnetic flux tube embedded in a magnetic field of $7$~G.  The temperature of the plasma flow is of the order of $10^5$~K while that of its environment is taken to be $2 \times 10^6$~K.  The electron number density of surrounding magnetized plasma has a typical for the TR/lower corona region value of $2 \times 10^{9}$~cm$^{-3}$.  Under these conditions, the Alfv\'en speed inside the jet is equal to $78.3$~km\,s$^{-1}$.  We model the surge as a moving magnetic flux tube for two magnetic field configurations: (i) a twisted tube surrounded by plasma with homogeneous background magnetic field, and (ii) a twisted tube which environment is plasma with also twisted magnetic field.  The magnetic field twist in given region is characterized by the ratio of azimuthal to the axial magnetic field components evaluated at the flux tube radius.  The numerical studies of appropriate dispersion relations of MHD modes supported by the plasma flow in both magnetic field configurations show that unstable against Kelvin--Helmholtz instability can only be the MHD waves with high negative mode numbers and the instability occurs at sub-Alfv\'enic critical flow velocities in the range of $25$--$50$~km\,s$^{-1}$.
\end{abstract}


\end{frontmatter}


\section{Introduction}
\label{sec:intro}
Surges are phenomena in which dark dense mass are ejected in the solar atmosphere from chromospheric into coronal heights. Usually, they appear as straight or slightly curved ejective structures, and they often recur \citep{roy1973a,roy1973b,svestka1976,tandberg1977,foukal1990}.  At first, they were studied in H$\alpha$ by \cite{newton1942} and by \cite{mcmath1948}.  They have a typical size of $38\,000$--$220\,000$~km, a transverse velocity of $30$--$200$~km\,s$^{-1}$, and a lifetime of $10$--$20$~min.  The rotational or helical motions were, on occasions, also observed in surge activity
\citep{gu1994,canfield1996}.  Surges were also observed as emission in He {\footnotesize\textsc{II}} $304$~\AA\ solar images \citep{bohlin1975,georgakilas1999}, obtained by the slitless XUV spectrograph on a \emph{Skylab\/} mission and with the Extreme Ultraviolet Imaging Telescope on board the \emph{Solar and Heliospheric Observatory\/} (\emph{SOHO}), respectively.  With developing observational instruments and spacecrafts, chromospheric ejections were detected by using data from the $50$~cm Swedish Vacuum Solar Telescope and the \emph{Transition Region and Coronal Explorer\/} (\emph{TRACE}, \citealp{brooks2007}), as well as from the Big Bear Solar Observatory \citep{chen2008} and with the Solar Ultraviolet Measurements of Emitted Radiation (SUMER) spectroscopic observations on board \emph{SOHO} \citep{chenpf2008}.

Based on observational studies, it is now accepted that the driving mechanism of mass ejection in surges is magnetic reconnection at chromospheric heights. \cite{kurokawa1993} found that H$\alpha$ surges occur at the very beginning phases of magnetic-flux emergence, and suggested that surges are produced by magnetic reconnection between the emerging flux and the pre-existing magnetic flux.  \cite{canfield1996} reported that circumstances favorable to magnetic reconnection are produced by moving satellite spots in a surge-productive region.   \cite{uddin2012} presented a multi-wavelength study of recurrent surges originated due to the photospheric reconnections.  Very recently, \cite{chandra2015} reported a multi-wavelength study of solar jets on 2010 December 11 using the \emph{Solar Dynamics Observatory\/} (\emph{SDO\/}, \citealp{pesnell2012}) data.  They found an increase in the amplitude of oscillations close to their footpoints of the observed jets, which provides the evidence for the wave-induced reconnection as a mechanism for jets triggering.  \cite{canfield1996} also showed that a high-temperature X-ray jet and a cool untwisted surge can coexist located side by side at the site (see Fig.~10c in their paper).  \cite{shibata1992} and \cite{yokoyama1995} succeeded to reproduce surge mass ejections from chromospheric heights by magnetic reconnection between an emerging flux and a pre-existing magnetic field.  This numerical result was later confirmed by \cite{nindos1998} who studied the radio properties of $18$ X-ray coronal jets as observed by the \emph{Yohkoh\/} SXT \citep{tsuneta1991} using Nobeyama Radioheliograph $17$~GHz data.  From the SXT images, \cite{nindos1998} computed the coronal plasma parameters at the location of the surge.  At the time of maximum surge activity, they found electron temperature $T_{\rm e} = 2.8 \times 10^6$~K and emission measure $EM = 5.0 \times 10^{45}$~cm$^{-3}$, as well as derived constraints on the ejecta electron number density, notably $n_{\rm e} < 6.1 \times 10^{10}$~cm$^{-3}$.  H$\alpha$ surges and associated soft X-ray loops were also studied by \cite{schmieder1994} who performed simultaneous observations of NOAA AR 6850 on 1991 October 7, made with the MSDP spectrograph operating on the solar tower in Meudon and with the \emph{Yohkoh\/} SXT.  By measuring the volume emission measures of the two flaring loops (northern and southern ones) and the surge region (mid-part of the surge), \cite{schmieder1994} concluded that at 10:24--10:30 UT the temperature was ($3$--$4) \times 10^6$~K and the volume emission measure was $10^{47}$~cm$^{-3}$.  Assuming a volume of ($3$--$10) \times 10^{27}$~cm$^3$, they derived an electron number density $n_{\rm e} = (3$--$6) \times 10^9$~cm$^{-3}$.  \cite{kayshap2013} have observed a solar surge in NOAA AR 11271 using the \emph{SDO\/} data on 2011 August 25, possibly triggered by chromospheric activity.  They also measured the temperature and density distribution of the observed surge during its maximum rise and found an average temperature and a number density of $2 \times 10^6$~K and $4.17 \times 10^9$~cm$^{-3}$, respectively.  \cite{brigitte1996} studied the conditions for flares and surges in AR 2744 on 1980 October 21 and 22 using observations from the \emph{Solar Maximum Mission\/} satellite and coordinated ground-based observations, which together covered a wide temperature range from ${<}10^4$~K to ${>}10^7$~K.  In particular, the detected surge on October 22 had a total emission measure of $4.9 \times 10^{44}$~cm$^{-3}$ and duration of about $2000$~s.  The rough estimations of temperature and electron number density yielded $T_{\rm e} \sim 10^4$~K and $n_{\rm e} \sim 10^{12}$~cm$^{-3}$.  Similar values for the temperature and electron number density of H$\alpha$ surges were obtained earlier by \cite{jain1987} from observations of a surge prominence in AR 17212 on 1980 October 30 made at an interval of $5$~s and $10$~s in the H$\alpha$ line center, through a Halle filter of $0.7$~\AA\ passband in conjunction with a $15$~cm aperture solar spar telescope.

As seen, the electron number density and the temperature of solar surges can vary in rather wide limits, from ${\sim}10^{12}$~cm$^{-3}$ and $10^4$~K for cool H$\alpha$ surges to ${\sim}10^{9}$~cm$^{-3}$ and $10^6$~K for high-temperature EUV surges.  Since each surge is a jet in a well-defined magnetic flux tube, it is naturally to expect that the magnetohydrodynamic waves propagating along the magnetized plasma flow can become unstable against the Kelvin--Helmholtz (KH) instability.  It is well-known that KH instabilities occur when two fluids of different densities or different speeds flow by each other.  In the solar atmosphere, which is made of a very hot and practically fully ionized plasma, the two flows come from an expanse of plasma erupting off the Sun's surface as it passes by plasma that is not erupting.  The difference in flow speeds and densities across this boundary sparks the instability that builds into the waves.  When the instability reaches its nonlinear stage, vortices might form, reconnection might be initiated and plasma structures might detach.  Now, after the launch of \emph{SDO\/} satellite, due to its high spatial and temporal resolution, the KH instability of the coronal mass ejection reconnection outflow layer in the lower corona, occurred on 2010 November 3, has been imaged by \cite{foullon2013}.  Very recently, that KH instability observation was modeled by \cite{Zhelyazkov2014b}.  A concise but very good exploration of KH instabilities in the solar atmosphere in view of their interpretation from observations the reader can find in \cite{taroyan2011}.  The aim of this study is to see whether MHD waves traveling along the surge jet can become unstable within its velocity range of $20$--$200$~km\,s$^{-1}$.  In the following Sect.~\ref{sec:model}, we will build up simplified models for the surge.  Next Sect.~\ref{sec:dispeqn} deals with the derivation of the MHD wave dispersion relations, while in Sect.~\ref{sec:numerics} we will numerically analyze the dependence of the linear/threshold KH instability on relevant physical parameters of the surge and its environment.  The last Sect.~\ref{sec:concl} summarizes our results.

\section{Surge models, basic parameters, and governing equations}
\label{sec:model}
We explore one of the four H$\alpha$ surges observed by \cite{brooks2007} in the solar active region NOAA AR 8227 (N$26^{\circ}$, E$09^{\circ}$) on 1998 May 30 from 7:50 to 16:50 UT.  The electron number density derived from the emission measure analysis of the \emph{TRACE\/} Fe {\footnotesize\textsc{IX}} $171$~\AA\ images is equal to $n_{\rm i} = 3.83 \times 10^{10}$~cm$^{-3}$ (the label `i' stands for \emph{interior}), and the jet velocity is of the order of $45$--$50$~km\,s$^{-1}$.  Estimated chromospheric magnetic field is $B_{\rm foot} = 25$~G and from the conservation of magnetic flux between the reconnection region and the photosphere one finds that the coronal magnetic field is $7$--$10$~G.  \cite{brooks2007} claim that their observations confirm the emerging flux regions model of \cite{kurokawa1993} and \cite{shibata1994}; moreover the clear evidence of spatiotemporal correlations between chromospheric (Ca {\footnotesize\textsc{II}} K, H$\alpha$) and coronal brightenings (\emph{TRACE\/} Fe {\footnotesize\textsc{IX}} $171$~\AA) indicates that the chromosphere is heated up to coronal temperatures at the surge footpoint by the energy release during magnetic field reconnection, as shown in the numerical simulation of \cite{yokoyama1996}.

It is clear that the main body of the aforementioned surge is positioned at the TR/lower corona region.  Thus, we can take the electron number density of the surge environment to be equal to $n_{\rm e} = 2 \times 10^9$~cm$^{-3}$ (the label `e' stands for \emph{exterior}), and its temperature reasonably can be $T_{\rm e} = 2 \times 10^6$~K.  Concerning the surge temperature, we suppose it to be equal to $T_{\rm i} = 10^5$~K.  Then with a background magnetic field $B_{\rm e} = 7$~G and a density contrast $\eta = \rho_{\rm e}/\rho_{\rm i} = 0.052$ (we assume that plasma densities in both media are homogeneous), we have the following characteristic sound and Alfv\'en speed inside the surge and in the surrounding magnetized plasma: $c_{\rm si} = 37$~km\,s$^{-1}$, $v_{\rm Ai} \cong 78.3$~km\,s$^{-1}$ (more exactly $78.314$~km\,s$^{-1}$, which value determines the magnetic field inside the jet to be equal to $B_{\rm i} = 7.04$~G), and $c_{\rm se} \cong 196$~km\,s$^{-1}$, $v_{\rm Ae} = 341$~km\,s$^{-1}$.  Thus, the plasma betas of the two media are respectively $\beta_{\rm i} = 0.27$ and $\beta_{\rm e} = 0.28$.  Incompressible plasma is a good approximation to study the KH instability, though the observed values are not favorable for the incompressibility. Nevertheless, in the following we consider the incompressible plasma inside and outside the surge.

\begin{figure}[t]
   \centering
   \includegraphics[height=.30\textheight]{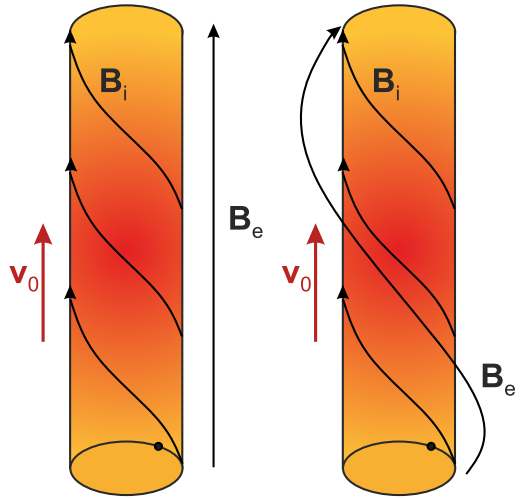}
   \caption{Equilibrium magnetic field geometries of a H$\alpha$ solar surge.}
   \label{fig:fig1}
\end{figure}

We model the surge as a vertically moving with a velocity $\boldsymbol{v}_0$ cylindrical flux tube with radius $a = \Delta \ell/2$ (see Fig.~\ref{fig:fig1}), where $\Delta \ell = 7$~Mm is the surge width.  Our frame of reference is attached to the TR/coronal plasma that implies that $\boldsymbol{v}_0$ is the relative jet velocity with respect to its environment.  We must mention that because the density contrast, $\eta$, is relatively high, in such a case, like in spicules, the occurrence of a KH instability, for instance of kink ($m = 1$) waves, becomes possible at generally high Alfv\'en Mach numbers (the Alfv\'en Mach number is defined as the ratio of jet velocity to Alfv\'en speed inside the jet, $M_{\rm A} = v_0/v_{\rm Ai}$) and correspondingly at high critical flow velocities being far beyond the speeds accessible for surges/spicules in the solar atmosphere \citep{zhelyazkov2012,ivan2012}.  This circumstance implies that the only possible way for emerging a KH instability in surges is the excitation of higher MHD harmonics that can become unstable at sub-Alfv\'enic flow velocities in twisted tubes \citep{zaqarashvili2010}.  Here, we consider two possible magnetic field geometries: (i) a moving twisted magnetic flux tube embedded in untwisted magnetic field $\boldsymbol{B}_{\rm e}$ (the left tube in Fig.~\ref{fig:fig1}), and (ii) a moving twisted magnetic flux tube surrounded by plasma with twisted magnetic field lines (the right tube in Fig.~\ref{fig:fig1}).  In our cylindrical coordinate system ($r,\varphi,z$) the magnetic field has the following form: $\boldsymbol{B} = \left(0, B_\varphi(r),B_z(r)\right)$, and the flow profile inside the tube is $\boldsymbol{v}_0 = (0,0,v_0)$.  In general, $v_0$ can be a function of $r$, but we consider the simplest homogeneous case.  The unperturbed magnetic field $\boldsymbol{B}$ and the pressure $p$ satisfy the pressure balance equation

\begin{equation}
\label{eq:equilib}
    \frac{\mathrm{d}}{\mathrm{d}r}\left( p + \frac{B_{\varphi}^2 + B_{z}^2}{2\mu}
    \right) = -\frac{B_{\varphi}^2}{\mu r},
\end{equation}
where $\mu$ is the magnetic permeability.

As the unperturbed parameters depend on the $r$ coordinate only, the perturbations can be Fourier analyzed with $\exp[-\mathrm{i}(\omega t - m \varphi - k_z z)]$.  The equations governing the incompressible plasma dynamics  are \citep{goossens1992}
\begin{eqnarray}
    \frac{\mathrm{d}^2 p_{\rm tot}}{\mathrm{d}r^2} + \left[ \frac{C_3}{rD} \frac{\mathrm{d}}{\mathrm{d}r} \left( \frac{rD}{C_3} \right) \right] \frac{\mathrm{d}p_{\rm tot}}{\mathrm{d}r} \nonumber \\
    \nonumber \\
    {}+ \left[ \frac{C_3}{rD} \frac{\mathrm{d}}{\mathrm{d}r} \left( \frac{rC_1}{C_3} \right) + \frac{1}{D^2} \left(  C_2 C_3 - C_1^2 \right) \right]p_{\rm tot} = 0.
\label{eq:ode}
\end{eqnarray}
where
\[
    D = \rho \left( \Omega^2 - \omega_{\rm A}^2 \right), \quad C_1 = -\frac{2mB_{\varphi}}{\mu r^2}\left( \frac{m}{r}B_{\varphi} + k_z B_z \right),
\]
\[
     C_2 = - \left( \frac{m^2}{r^2} + k_z^2 \right),
\]
\[
     C_3 = D^2 + D\frac{2 B_{\varphi}}{\mu}\frac{\mathrm{d}}{\mathrm{d} r} \left( \frac{B_{\varphi}}{r} \right) - \frac{4 B_{\varphi}^2}{\mu r^2}\rho \omega_{\rm A}^2,
\]
\begin{equation}
\label{eq:omegaa}
    \omega_{\rm A} = \frac{\boldsymbol{k}\cdot \boldsymbol{B}}{\sqrt{\mu \rho}} = \frac{1}{\sqrt{\mu \rho}}\left( \frac{m}{r}B_\varphi + k_z B_z\right)
\end{equation}
is the local Alfv\'en frequency,
\begin{equation}
\label{eq:omegaD}
    \Omega = \omega - \boldsymbol{k}\cdot \boldsymbol{v}_0
\end{equation}
is the Doppler-shifted frequency, and $p_{\rm tot}$ is total (thermal $+$ magnetic) pressure perturbation.  Radial displacement $\xi_r$ is expressed through the total pressure perturbation as
\begin{equation}
\label{eq:displ}
    \xi_{r} = \frac{D}{C_3}\frac{\mathrm{d} p_{\rm tot}}{\mathrm{d} r} + \frac{C_1}{C_3}p_{\rm tot}.
\end{equation}
The solution to this equation depends on the magnetic field and density profile.  To obtain the dispersion relation of MHD modes, we find the solutions to Eqs.~(\ref{eq:ode}) and (\ref{eq:displ}) inside and outside the tube and merge the solutions through boundary conditions.

\section{Wave dispersion relations}
\label{sec:dispeqn}
Before obtaining the wave dispersion relations for the two magnetic configurations, we first specify the twist of the magnetic flux tube to be a uniform one, that is,
\begin{equation}
\label{eq:magfbi}
    \boldsymbol{B}_{\rm i} = (0, Ar, B_{{\rm i}z}),
\end{equation}
where $A$ and $B_{{\rm i}z}$ are constant.  In the simpler case (left flux tube in Fig.~\ref{fig:fig1}) the magnetic field outside the tube is homogeneous, $\boldsymbol{B}_{\rm e} = (0, 0, B_{\rm e})$, and the solution to Eq.~(\ref{eq:ode}) inside the tube bounded at the tube axis is (see \citealp{zaqarashvili2014})
\begin{equation}
\label{eq:solution}
    p_{\rm tot}(r \leqslant a) = \alpha_{\rm i}I_m(\kappa_{\rm i}r),
\end{equation}
where $I_m$ is the modified Bessel function of order $m$ and $\alpha_{\rm i}$ is a constant.  Transverse displacement can be written using Eq.~(\ref{eq:displ}) as
\begin{eqnarray}
\label{eq:xir}
    \xi_{{\rm i}r} = \frac{\alpha_{\rm i}}{r}\left\{ \frac{\left( \Omega^2 - \omega_{\rm Ai}^2 \right)\kappa_{\rm i}r I_m^{\prime}(\kappa_{\rm i}r)}{\rho_{\rm i}\left( \Omega^2 - \omega_{\rm Ai}^2 \right)^2 - 4A^2 \omega_{\rm Ai}^2/\mu} \right. \nonumber \\
    \nonumber \\
    \left.
    - \frac{2mA\omega_{\rm Ai}I_m(\kappa_{\rm i}r)/\sqrt{\mu \rho_{\rm i}} }{\rho_{\rm i}\left( \Omega^2 - \omega_{\rm Ai}^2 \right)^2 - 4A^2 \omega_{\rm Ai}^2/\mu} \right\},
\end{eqnarray}
where the attenuation coefficient $\kappa_{\rm i}$ and Alfv\'en frequency $\omega_{\rm Ai}$ are given by
\begin{equation}
\label{eq:kappai}
	\kappa_{\rm i} = k_z\left[  1 - 4 A^2 \omega_{\rm Ai}^2/
        \mu \rho_{\rm i} \left( \Omega^2 -
        \omega_{\rm Ai}^2\right)^2 \right]^{1/2},
\end{equation}
\begin{equation}
\label{eq:omegaai}
    \omega_{\rm Ai} = \frac{mA + k_z B_{{\rm i}z}}{\sqrt{\mu \rho_{\rm i}}},
\end{equation}
and prime sign means a differentiation by the Bessel function argument.

The solution to Eq.~(\ref{eq:ode}) outside the flux tube bounded at infinity is
\begin{equation}
\label{eq:solution1}
    p_{\rm tot}(r > a) = \alpha_{\rm e}K_m(k_z r),
\end{equation}
where $K_m$ is the modified Bessel function of order $m$ and $\alpha_{\rm e}$ is a constant.  Transverse displacement can be written as
\begin{equation}
\label{eq:xier}
    \xi_{{\rm e}r} = \frac{\alpha_{\rm e}}{r}\frac{k_z r K_m^{\prime}(k_z r)}{\rho_{\rm e}\left( \omega^2 - \omega_{\rm Ae}^2 \right)},
\end{equation}
where, as before, the prime sign means a differentiation by the Bessel function argument, and the local Alfv\'en frequency is
\begin{equation}
\label{eq:OmegaAe}
    \omega_{\rm Ae} = \frac{k_z B_{\rm e}}{\sqrt{\mu \rho_{\rm e}}} = k_z v_{\rm Ae}.
\end{equation}
Here, $v_{\rm Ae} = B_{\rm e}/\sqrt{\mu \rho_{\rm e}}$ is the Alfv\'en speed in the tube environment.

The boundary conditions which merge the solutions inside and outside the twisted magnetic flux tube are the continuity of the radial component of the Lagrangian displacement
\begin{equation}
\label{eq:xicont}
    \xi_{{\rm i}r}|_{r=a} = \xi_{{\rm e}r}|_{r=a}
\end{equation}
and the total pressure perturbation \citep{bennett1999}
\begin{equation}
\label{eq:ptotcont}
    \left.p_{\rm tot\,i} - \frac{B_{{\rm i}\varphi}^2}{\mu a}\xi_{{\rm i}r}\right\vert_{r=a} = p_{\rm tot\,e}|_{r=a},
\end{equation}
where total pressure perturbations $p_{\rm tot\,i}$ and $p_{\rm tot\,e}$ are given by Eqs.~(\ref{eq:solution}) and (\ref{eq:solution1}).  Applying these boundary condition, after some algebra we finally derive the dispersion relation of the normal MHD modes propagating along a twisted magnetic flux tube with axial mass flow $\boldsymbol{v}_0$ surrounded by plasma embedded in a homogeneous magnetic field
\begin{eqnarray}
\label{eq:dispeq}
	\frac{\left( \Omega^2 -
    \omega_{\rm Ai}^2 \right)F_m(\kappa_{\rm i}a) - 2mA \omega_{\rm Ai}/\sqrt{\mu \rho_{\rm i}}}
    {\left( \Omega^2 -
    \omega_{\rm Ai}^2 \right)^2 - 4A^2\omega_{\rm Ai}^2/\mu \rho_{\rm i} } \nonumber \\
    \nonumber \\
    {} = \frac{P_m(k_z a)}
    {{\displaystyle \frac{\rho_{\rm e}}{\rho_{\rm i}}} \left( \omega^2 - \omega_{\rm Ae}^2
    \right) + A^2  P_m(k_z a)/\mu \rho_{\rm i}},
\end{eqnarray}
where, remember, $\Omega = \omega - \boldsymbol{k}\cdot \boldsymbol{v}_0$ is the Doppler-shifted wave frequency in the moving flux tube, and
\[
    F_m(\kappa_{\rm i}a) = \frac{\kappa_{\rm i}a I_m^{\prime}(\kappa_{\rm i}a)}{I_m(\kappa_{\rm i}a)} \quad \mbox{and} \quad P_m(k_z a) = \frac{k_z a K_m^{\prime}(k_z a)}{K_m(k_z a)}.
\]
A derivation of Eq.~(\ref{eq:dispeq}) starting from the basic equations of ideal magnetohydrodynamics the reader can see in \citealp{ivan2012}.

In the case when the outside magnetic field is also twisted (the right flux tube in Fig.~\ref{fig:fig1}), we consider that magnetic field, $\boldsymbol{B}_{\rm e}$, has the form \citep{zaqarashvili2014}
\begin{equation}
\label{eq:be}
    \boldsymbol{B}_{\rm e} = \left( 0, B_{{\rm e}\varphi} \frac{a}{r}, B_{{\rm e}z} \left( \frac{a}{r} \right)^2 \right),
\end{equation}
and the density is presented as $\rho = \rho_{\rm e}(a/r)^4$, so that the Alfv\'en frequency
\begin{equation}
\label{eq:omegaae}
    \omega_{\rm Ae} = \frac{m B_{{\rm e}\varphi} + k_z a B_{{\rm e}z}}{\sqrt{\mu \rho_{\rm e}}a}
\end{equation}
is constant, which allows us to find an analytical solution to the governing Eq.~(\ref{eq:ode}).  The total pressure perturbation outside the tube is governed by the Bessel-type equation
\begin{equation}
    \frac{\mathrm{d}^2 p_{\rm tot}}{\mathrm{d}r^2} + \frac{5}{r}\frac{\mathrm{d}p_{\rm tot}}{\mathrm{d}r} - \left( \frac{n^2}{r^2} + \kappa_{\rm e}^2 \right)p_{\rm tot} = 0,
\label{eq:besssel-like}
\end{equation}
where
\begin{equation}
\label{eq:n}
    n^2 = m^2 - \frac{4m^2 B_{{\rm e}\varphi}^2}{\mu \rho_{\rm e}a^2 \left( \omega^2 - \omega_{\rm Ae}^2 \right)} + \frac{8m B_{{\rm e}\varphi} \omega_{\rm Ae}}{\sqrt{\mu \rho_{\rm e}}a \left( \omega^2 - \omega_{\rm Ae}^2 \right)},
\end{equation}
and
\begin{equation}
\label{eq:kappae}
    \kappa_{\rm e}^2 = k_z^2 \left( 1 - \frac{4 B_{{\rm e}\varphi}^2 \omega^2}{\mu \rho_{\rm e} \left( \omega^2 - \omega_{\rm Ae}^2 \right)^2 a^2} \right).
\end{equation}
A solution to Eq.~(\ref{eq:besssel-like}) bounded at infinity is
\begin{equation}
\label{eq:ptote}
    p_{\rm tot}(r > a) = \alpha_{\rm e}\frac{a^2}{r^2}K_{\nu}(\kappa_{\rm e} r),
\end{equation}
where $\nu = \sqrt{4 + n^2}$ and $\alpha_{\rm e}$ is a constant.

Transversal displacement can be written as
\begin{eqnarray}
\label{eq:xie}
    \xi_{{\rm e}r} = \alpha_{\rm e} \frac{r \left( \omega^2 - \omega_{\rm Ae}^2 \right)\kappa_{\rm e}r K_{\nu}^{\prime}(\kappa_{\rm e} r)}{a^2 \rho_{\rm e}\left( \omega^2 - \omega_{\rm Ae}^2 \right)^2 - 4 B_{{\rm e}\varphi}^2 \omega^2/\mu} \nonumber \\
    \nonumber \\
    {}- \alpha_{\rm e}\frac{r}{a}\left\{ \frac{2a\left( \omega^2 - \omega_{\rm Ae}^2 \right)}{a^2 \rho_{\rm e}\left( \omega^2 - \omega_{\rm Ae}^2 \right)^2 - 4 B_{{\rm e}\varphi}^2 \omega^2/\mu} \right. \nonumber \\
    \nonumber \\
    \left.
    {}+ \frac{2m B_{{\rm e}\varphi} \omega_{\rm Ae}/\sqrt{\mu \rho_{\rm e}}}{a^2 \rho_{\rm e}\left( \omega^2 - \omega_{\rm Ae}^2 \right)^2 - 4B_{{\rm e}\varphi}^2 \omega^2/\mu} \right\}K_{\nu}(\kappa_{\rm e} r).
\end{eqnarray}

By applying boundary conditions (\ref{eq:xicont}) and (\ref{eq:ptotcont}), where the transversal displacements are given by Eqs.~(\ref{eq:xier}) and (\ref{eq:xie}), and total pressure perturbations $p_{\rm tot\,i}$ and $p_{\rm tot\,e}$, accordingly, by Eqs.~(\ref{eq:solution}) and (\ref{eq:ptote}), we obtain the dispersion relation of the normal MHD modes propagating along a twisted magnetic flux tube with axial mass flow $\boldsymbol{v}_0$ surrounded by plasma embedded in twisted magnetic field
\begin{eqnarray}
\label{eq:dispeqtwist}
	\frac{\left( \Omega^2 -
    \omega_{\rm Ai}^2 \right)F_m(\kappa_{\rm i}a) - 2mA \omega_{\rm Ai}/\sqrt{\mu \rho_{\rm i}}}
    {\left( \Omega^2 -
    \omega_{\rm Ai}^2 \right)^2 - 4A^2\omega_{\rm Ai}^2/\mu \rho_{\rm i} } \nonumber \\
    \nonumber \\
    {} = \frac{a^2 \left( \omega^2 - \omega_{\rm Ae}^2 \right)Q_\nu(\kappa_{\rm e} r) - G}
    {L - H \left[ a^2 \left( \omega^2 - \omega_{\rm Ae}^2 \right)Q_\nu(\kappa_{\rm e} r) - G \right]},
\end{eqnarray}
where
\[
    Q_\nu(\kappa_{\rm e} r) = \frac{\kappa_{\rm e}a K_{\nu}^{\prime}(\kappa_{\rm e} a)}{K_{\nu}(\kappa_{\rm e} a)}, \; L = a^2 \rho_{\rm e} \left( \omega^2 - \omega_{\rm Ae}^2 \right)^2 - \frac{4B_{{\rm e}\varphi}^2 \omega^2}{\mu},
\]

\[
    H = \frac{B_{{\rm e}\varphi}^2}{\mu a^2} - \frac{A^2}{\mu}, \quad G = 2a^2 \left( \omega^2 - \omega_{\rm Ae}^2 \right) + \frac{2ma B_{{\rm e}\varphi} \omega_{\rm Ae}}{\sqrt{\mu \rho_{\rm e}}}.
\]

Note that the left-hand sides of dispersion equations (\ref{eq:dispeq}) and (\ref{eq:dispeqtwist}) are identical (this is not surprising), but the right-hand sides are completely different due to the very different magnetic field environments.

\section{Numerical calculations and results}
\label{sec:numerics}
The main goal of our study is to determine under which conditions the MHD waves propagating along the jet can become unstable.  To conduct this investigation, it is necessary to assume that the wave frequency $\omega$ is a complex quantity, that is, $\omega \to \omega + \mathrm{i}\gamma$, where $\gamma$ is the instability growth rate, while the longitudinal wavenumber $k_z$ is a real variable in the wave dispersion relation.  Since the occurrence of the expected KH instability is determined primarily by the jet velocity, by searching for a critical or threshold value of it, we will gradually change its magnitude from zero to that critical value and beyond.  Thus, we have to solve the dispersion relations in complex variables obtaining the real and imaginary parts of the wave frequency, or as is commonly accepted, of the wave phase velocity $v_{\rm ph} = \omega/k_z$, as functions of $k_z$ at various values of the velocity shear between the surge and its environment, $v_0$.

Before starting the numerical job, we have to normalize all variables and to specified the input parameters.  The wave phase velocity, $v_{\rm ph}$, and the other speeds are normalized to the Alfv\'en speed inside the jet, $v_{\rm Ai}$, which is calculated on using the axial magnetic fields $B_{{\rm i}z}$.  The wavelength, $\lambda = 2\pi/k_z$, is normalized to the tube radius, $a$, that is equivalent to introducing a dimensionless wavenumber $k_z a$.  For normalizing the Alfv\'en frequency in the environment, $\omega_{\rm Ae}$, except the density contrast $\eta$ and the tube radius $a$, we have to additionally specify the ratio of the axial magnetic field components in both media, $b = B_{{\rm e}z}/B_{{\rm i}z}$.  For our surge and its environment that ratio is equal to $0.994373$ and because we consider the two plasmas as incompressible media, we shall take $b = 1$.  In the dimensionless analysis the flow speed, $v_0$, will be presented by the Alfv\'en Mach number $M_{\rm A} = v_0/v_{\rm Ai}$.

We first begin with the numerical solving Eq.~(\ref{eq:dispeq}).  As we already said, the kink ($m = 1$) mode might be in principal exited, but owing to the relatively high density contrast, $\eta = 0.052$, calculations show that this mode can become unstable at Alfv\'en Mach numbers larger than $6.36$, which means a critical jet velocity of $498$~km\,s$^{-1}$, which is inaccessible for solar surges.  The excitations of unstable MHD waves with higher positive mode numbers $m = 2$ or $m = 3$ also requires very high critical plasma flow speeds beyond the upper limit of $200$~km\,s$^{-1}$ for surges.  A distinctive decrease of the instability critical Alfv\'en Mach number/jet speed one can achieve at the propagation of MHD waves with negative mode numbers.
\begin{figure}[!ht]
\centering
    \includegraphics[width=7.5cm]{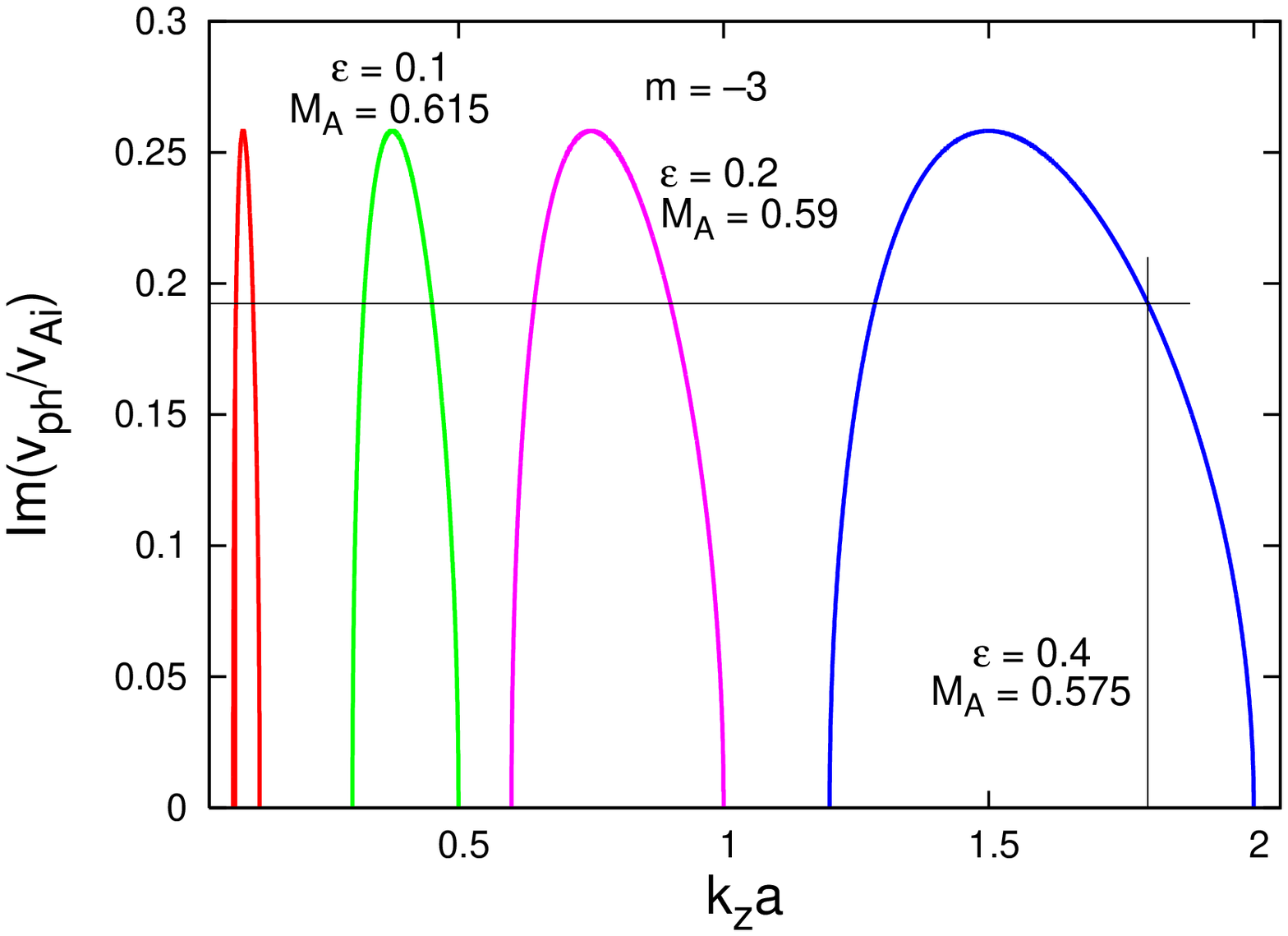} \\

\vspace{1mm}
    \includegraphics[width=7.5cm]{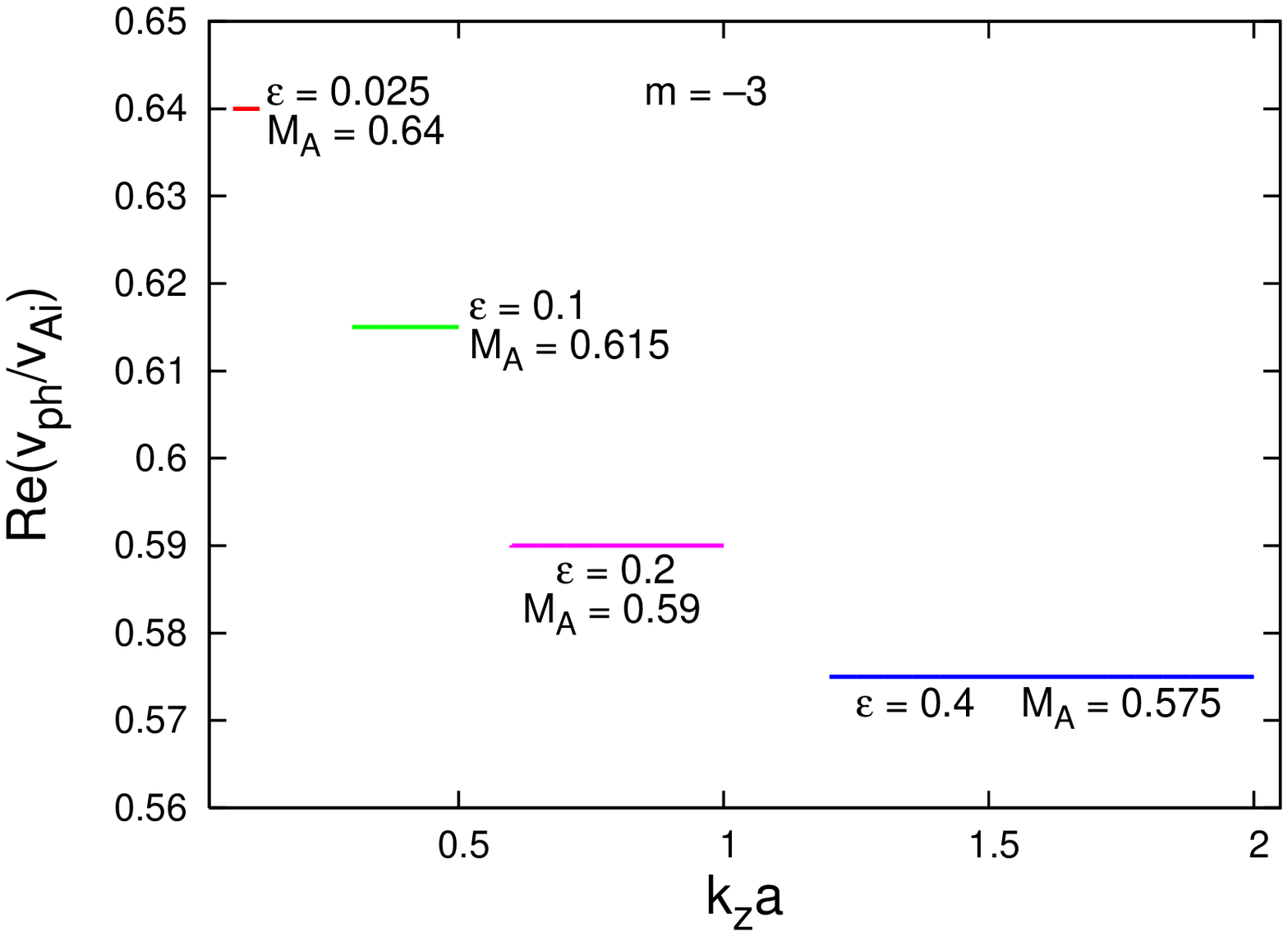} \\
   \caption{(\emph{Top panel}) Growth rates of the unstable $m = -3$ MHD mode propagating on incompressible jets in four different twisted internal magnetic fields (with $\varepsilon = 0.052$, $0.1$, $0.2$, and $0.4$) at $\eta = 0.052$, $b = 1$, and corresponding critical Alfv\'en Mach numbers.  For $k_za = 1.8$ the wavelength of the unstable $m = -3$ harmonic is $\lambda_{\rm KH} = 12.2$~Mm, and the wave growth rate is $\gamma_{\rm KH} = 0.008$~s$^{-1}$.  (\emph{Bottom panel}) Marginal dispersion curves of the unstable $m = -3$ MHD mode for the critical Alfv\'en Mach numbers as functions of the magnetic field twist parameter $\varepsilon$.  At $k_z a = 1.8$ the critical jet velocity is $v_0^{\rm cr} = 45$~km\,s$^{-1}$.}
   \label{fig:fig2}
\end{figure}
\begin{figure}[!ht]
\centering
    \includegraphics[width=7.5cm]{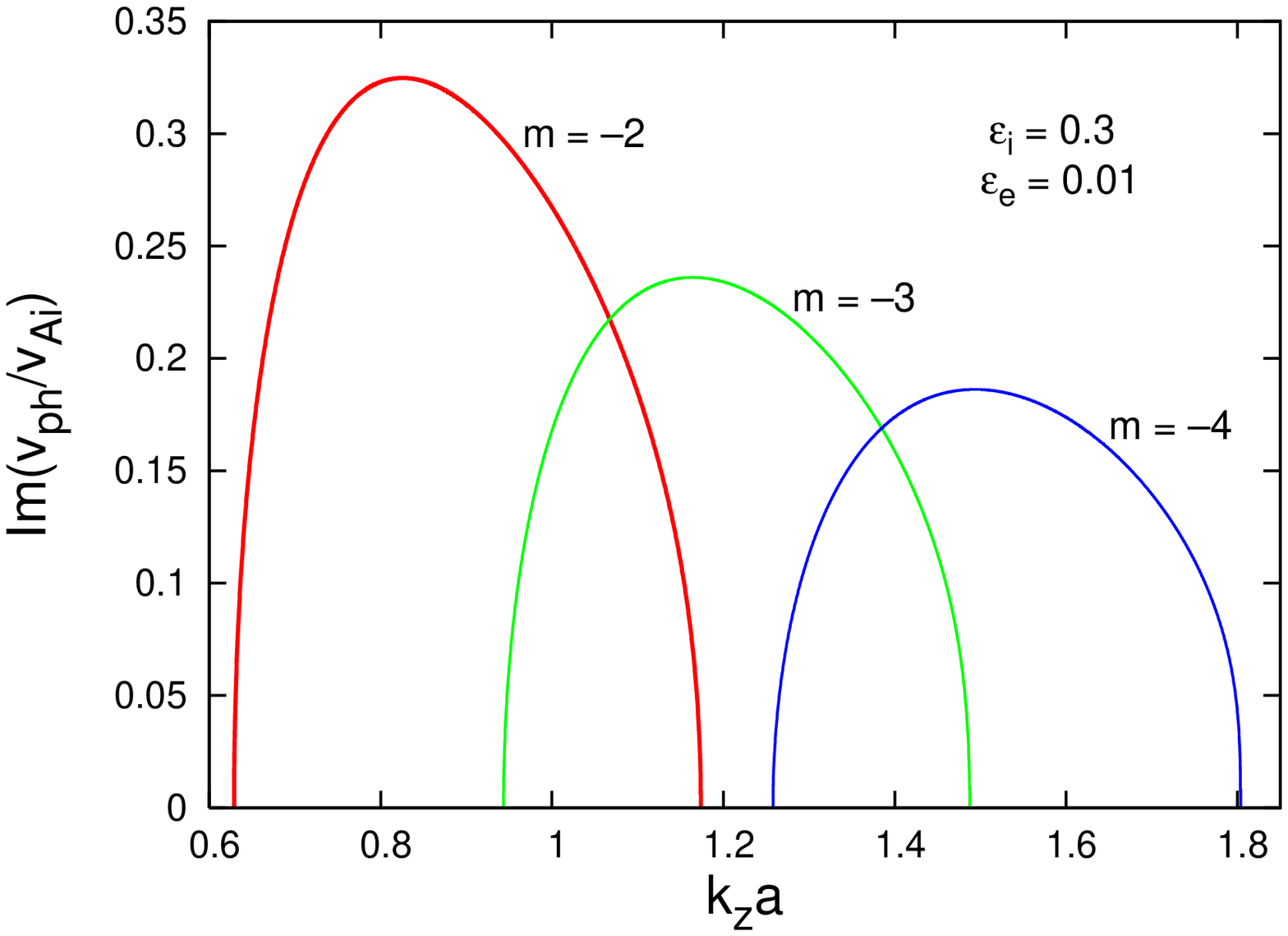} \\

\vspace{1mm}
    \includegraphics[width=7.5cm]{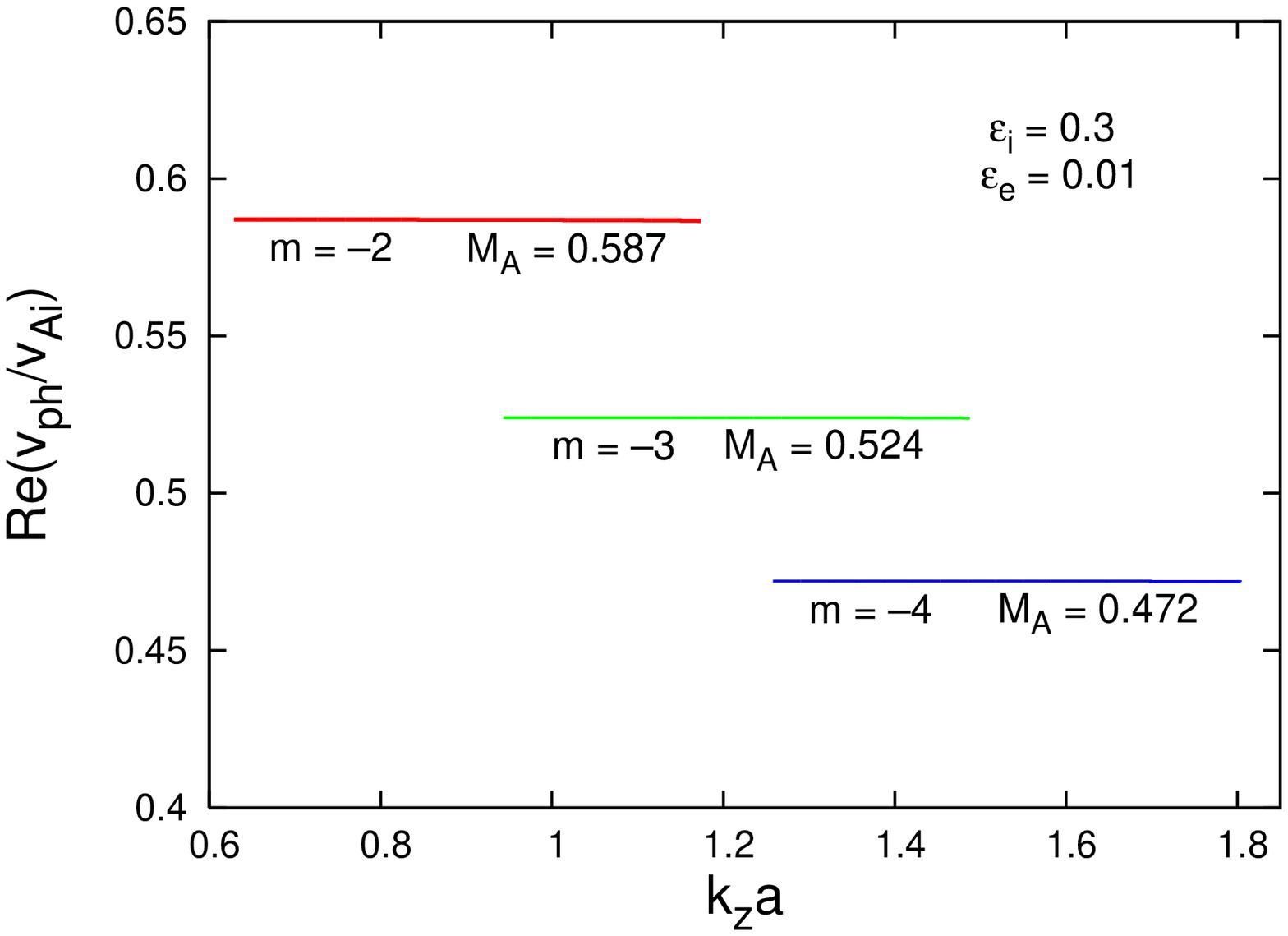} \\
   \caption{(\emph{Top panel}) Growth rates of the unstable $m = -2$, $m = -3$, and $m = -4$ MHD modes propagating on incompressible twisted jets with $\varepsilon_{\rm i} = 0.3$ in a twisted external magnetic field with $\varepsilon_{\rm e} = 0.01$ at $\eta = 0.052$, $b = 1$, and critical Alfv\'en Mach numbers equal to $0.587$, $0524$, and $0.472$.  (\emph{Bottom panel}) Marginal dispersion curves of the unstable $m = -2$, $m = -3$, and $m = -4$ MHD modes for the critical Alfv\'en Mach numbers at the magnetic fields twist parameters $\varepsilon_{\rm i} = 0.3$ and $\varepsilon_{\rm e} = 0.01$.  The critical surge velocities of these modes are correspondingly equal to $46$~km\,s$^{-1}$, $41$~km\,s$^{-1}$, and $37$~km\,s$^{-1}$.}
   \label{fig:fig3}
\end{figure}
\begin{figure}[!ht]
\centering
    \includegraphics[width=7.5cm]{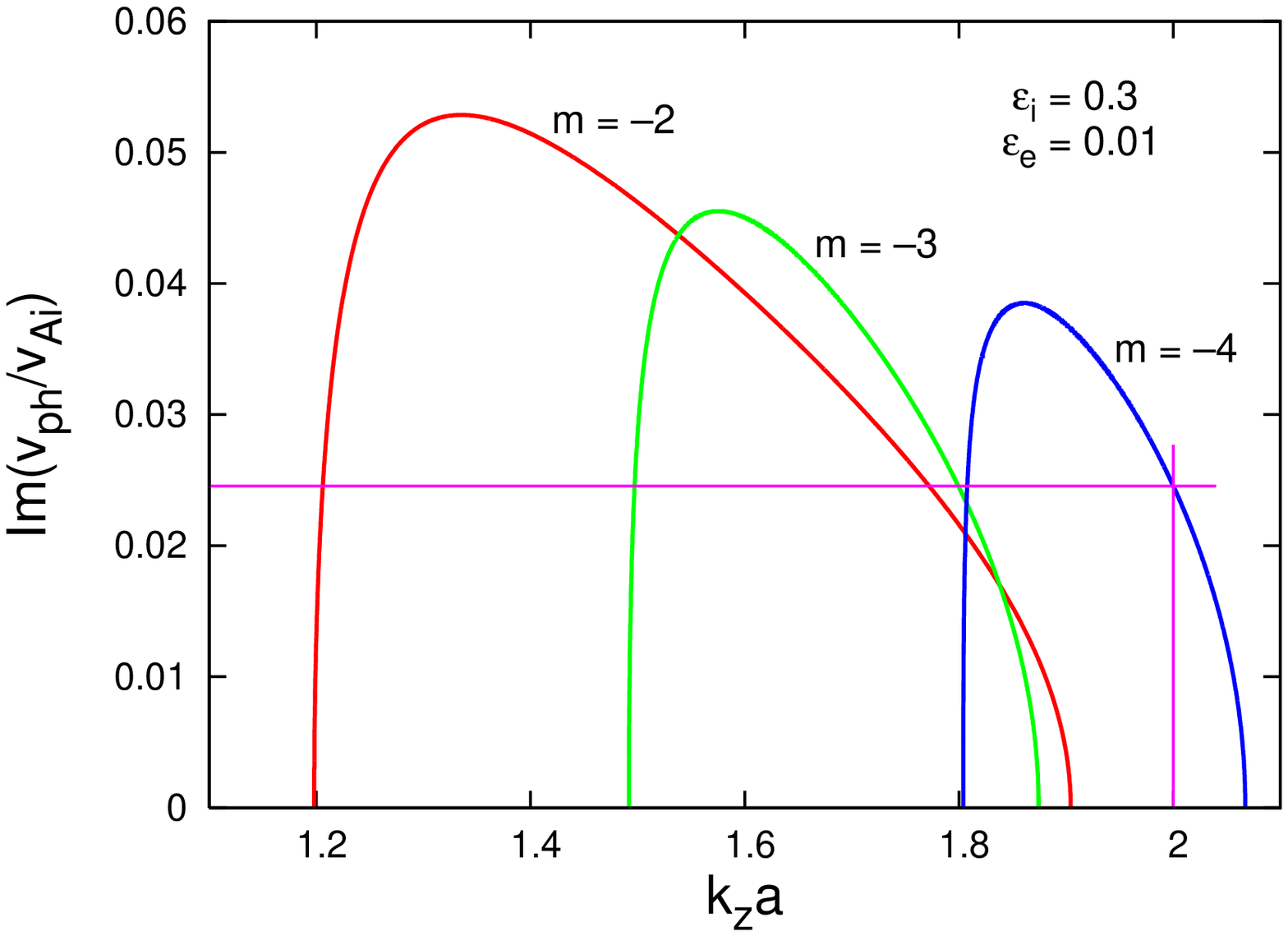} \\

\vspace{1mm}
    \includegraphics[width=7.5cm]{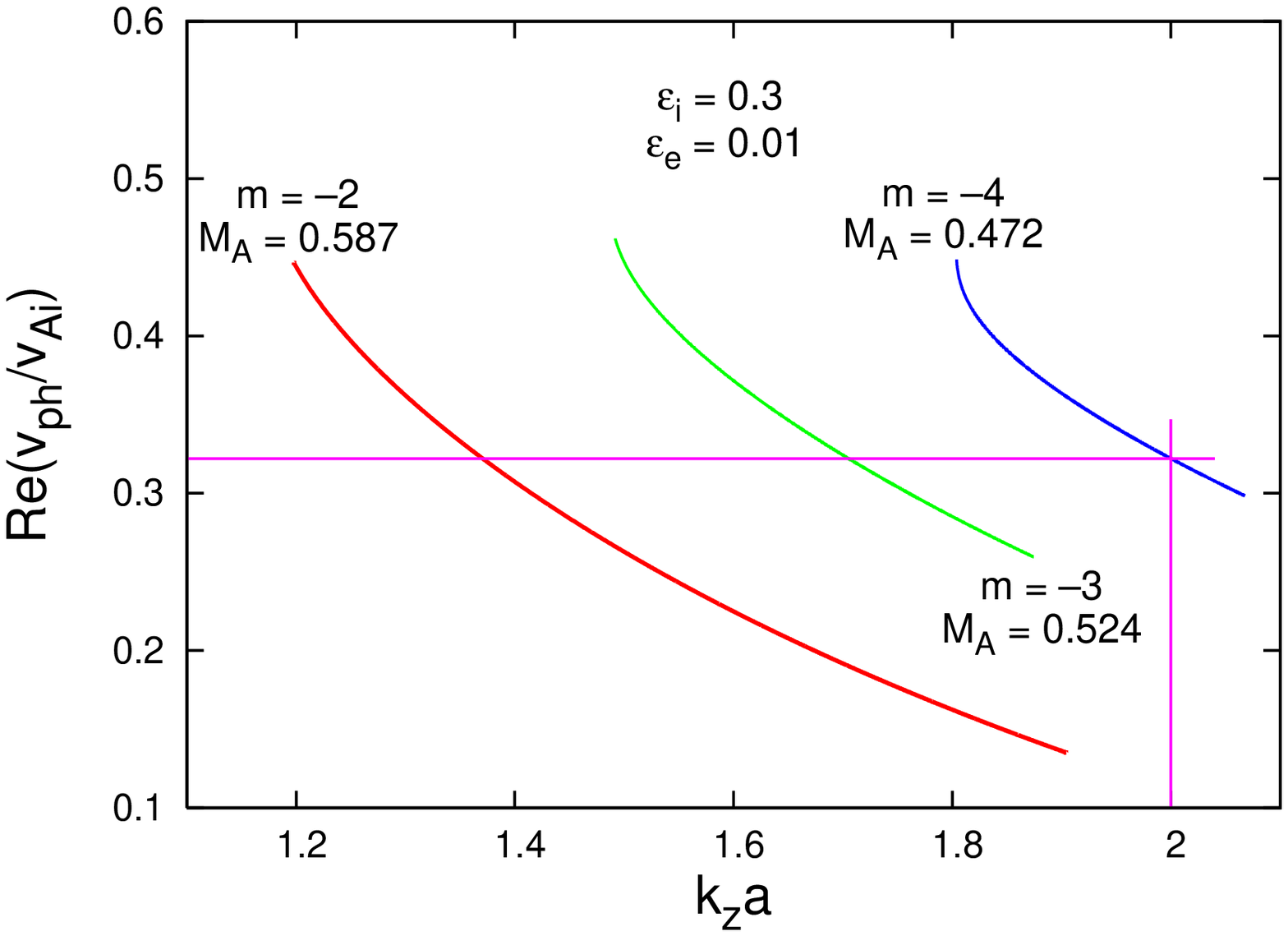} \\
   \caption{(\emph{Top panel}) Growth rates of the unstable $m = -2$, $m = -3$, and $m = -4$ MHD modes propagating on incompressible twisted jets with $\varepsilon_{\rm i} = 0.3$ in a twisted external magnetic field with $\varepsilon_{\rm e} = 0.01$ at $\eta = 0.052$, $b = 1$, and corresponding critical Alfv\'en Mach numbers.  For $k_za = 2$ the wavelength of the unstable $m = -4$ harmonic is $\lambda_{\rm KH} = 11$~Mm, and the wave growth rate is $\gamma_{\rm KH} = 0.001$~s$^{-1}$.  (\emph{Bottom panel}) Marginal dispersion curves of the unstable $m = -2$, $m = -3$, and $m = -4$ MHD modes for the critical Alfv\'en Mach numbers at the magnetic fields twist parameters $\varepsilon_{\rm i} = 0.3$ and $\varepsilon_{\rm e} = 0.01$.  The critical surge velocity of the $m = -4$ harmonic at $k_z a = 2$ is equal to $25.2$~km\,s$^{-1}$.  By contrast to the instability windows plotted in the bottom panel of Fig.~\ref{fig:fig3}, in this family of instability windows the normalized phase velocities are not constant.}
   \label{fig:fig4}
\end{figure}
Numerical calculations of Eq.~(\ref{eq:dispeq}) for the $m = -3$ MHD mode give in four instability windows on the $k_za$-axis whose position and width depends upon the magnetic field twist parameter $\varepsilon$ (see Fig.~\ref{fig:fig2}).  The critical jet speeds for emergence of a KH instability accordingly are $50.1$~km\,s$^{-1}$, $48.2$~km\,s$^{-1}$, $46.2$~km\,s$^{-1}$, and $45$~km\,s$^{-1}$.  The two narrow windows corresponding to $\varepsilon = 0.025$ and $\varepsilon = 0.1$ are practically inapplicable to our surge--environment configuration: the wavelength, $\lambda = \pi \Delta \ell/k_z a$, of unstable $m = -3$ harmonics becomes comparable to the surge's height; for instance, for $k_z a = 0.4$ (the middle of the second instability window) the wavelength of the unstable $m = -3$ harmonic at $\varepsilon = 0.1$ is $\lambda_{\rm KH} = 55$~Mm.  Actually only in the forth instability window (at $\varepsilon = 0.4$) one can have real unstable $m = -3$ MHD mode: for instance, at $k_za = 1.8$ the wavelength is $\lambda_{\rm KH} = 12.2$~Mm, and the corresponding dimensionless wave phase velocity growth rate is equal to $0.19245$ which implies a wave growth rate $\gamma_{\rm KH} = 0.008$~s$^{-1}$.  If we shift to the left at $k_z a = 1.5$ (with maximal normalized wave phase velocity growth rate), the wavelength is $\lambda_{\rm KH} = 14.7$~Mm, and the corresponding wave growth rate is a little bit higher, namely $\gamma_{\rm KH} = 0.0085$~s$^{-1}$.  The numerical study of the fluting-like harmonic $m = -2$ yields similar results; the only difference is that the fourth instability window locks at $k_z a = 1.6$.  Acceptable KH instability wavelengths and growth rates one can get again for $\varepsilon = 0.4$.  The critical jet velocities for the $m = -2$ mode, however, are a little bit higher---they are in the range of $55$--$60$~km\,s$^{-1}$.  Note that normalized wave phase velocity on given dispersion curve in the bottom panels of Figs.~\ref{fig:fig2} and \ref{fig:fig3} is equal to its label $\mathsf{M_{\sf A}}$.  Therefore, the unstable perturbations are frozen in the flow and consequently they are vortices rather than waves.  This observation is consistent with the KH instability in the hydrodynamics that deals with unstable vortices.

A much more interesting picture one obtains when study the more complicated case of a twisted magnetic flux tube surrounded by plasma embedded in a twisted background magnetic field.  Our choice for the twist characteristics of the two magnetic fields (internal and external ones) are $\varepsilon_{\rm i} = 0.3$ and $\varepsilon_{\rm e} = 0.01$, correspondingly.  Numerical solving Eq.~(\ref{eq:dispeqtwist}) for the three mode numbers $m = -2$, $m = -3$, and $m = -4$ gives for each mode number two instability windows: one of them with relatively high maximal growth rate, and a second window, next to the former, with one order lower maximal growth rate.  These two families of instability windows, for clarity, are presented in separate figures, Figs.~\ref{fig:fig3} and \ref{fig:fig4}.  As seen from Fig.~\ref{fig:fig3}, a real KH instability one can observe mostly for the $m = -4$ MHD mode and partly for the $m = -3$ harmonic.  The wave growth rates of unstable modes are of the same order like those illustrated in Fig.~\ref{fig:fig2}, i.e., few inverse milliseconds.  The critical flow velocities for the $m = -2$, $m = -3$, and $m = -4$ modes are correspondingly equal to $46$~km\,s$^{-1}$, $41$~km\,s$^{-1}$, and $37$~km\,s$^{-1}$.  The second family of instability windows, shown in Fig.~\ref{fig:fig4}, has two distinct peculiarities: first, the instability windows are shifted to the right-hand side of the $k_z a$-axis, i.e., the propagation range of unstable MHD modes is extended, and second, the marginal dispersion curves are not constant---the normalized wave phase velocities gradually decrease with increasing the dimensionless wavenumber.  If we fix the normalized wavenumber to be $k_z a = 2$, the wavelength of the unstable $m = -4$ mode is $\lambda_{\rm KH} = 11$~Mm, and at Im$(v_{\rm ph}/v_{\rm Ai}) = 0.02456$ the wave growth rate is $\gamma_{\rm KH} = 0.001$~s$^{-1}$, much lower than the observable growth rates in the first family of instability windows.  Note also, that the critical jet velocity for emerging a KH instability now is remarkably lower---its normalized value is $0.322$ that implies $v_0^{\rm cr} = 25.2$~km\,s$^{-1}$, i.e., the half of the surge speed evaluated by \cite{brooks2007}.  Thus, one can conclude that high-harmonic MHD modes can become unstable against the KH instability for accessible sub-Alfv\'enic velocities---this is more pronounced in the case when both magnetic fields are twisted.

\section{Discussion and conclusion}
\label{sec:concl}
In this paper, we have studied the condition under which MHD modes traveling on a H$\alpha$ solar surge can become unstable against the Kelvin--Helmholtz instability.  Our model for the surge is a vertically moving twisted magnetic cylindrical flux tube that might be surrounded by plasma embedded in homogeneous magnetic field or by magnetized plasma with twisted magnetic field.  In each case the twist of given magnetic field (internal or external one) is characterized by the ratio of azimuthal magnetic field component at the inner surface of the tube to its longitudinal component.  Among the MHD modes which propagate along the moving tube only the high harmonics with negative mode numbers can become unstable at accessible critical flow velocities.  One can observe practically a developing KH instability if only the parameter $\varepsilon$ characterizing the magnetic field twist is large enough.  For instance, at the first magnetic field configuration (the left flux tube in Fig.~\ref{fig:fig1}) the $m = -3$ harmonic becomes unstable at critical flow velocity of $45$~km\,s$^{-1}$ and wavelength $\lambda_{\rm KH} = 12.2$~Mm with a linear growth rate $\gamma_{\rm KH} = 0.008$~s$^{-1}$ as the magnetic field twist parameter $\varepsilon = 0.4$.  We note, that exploring the KH instability \citep{zhelyazkov2014a} in a high-temperature solar surge, like that observed by \cite{kayshap2013}, for the same mode at the same position on the $k_z a$-axis and the same value of $\varepsilon$, one obtains a wave growth rate of $0.033$~s$^{-1}$ being exactly equal to the growth rate of the imaged KH instability in a coronal mass ejecta in the lower corona by \cite{foullon2013}.  On the other hand, the aforementioned $\gamma_{\rm KH} = 0.008$~s$^{-1}$ is of the same order as the growth rate of $0.003$~s$^{-1}$ of vortex-shaped features along the interface between an erupting (dimming) region and the surrounding corona imaged by the \emph{SDO}/AIA as reported by \cite{ofman2011}.  All these comparisons allows us to believe that the KH instability might be imaged in surges, too.  The second magnetic field configuration (the right flux tube in Fig.~\ref{fig:fig1}), reveals some new aspects of the KH instability, notably for a fixed pair of magnetic fields twist parameters, in our case equal to $\varepsilon_{\rm i} = 0.3$ and $\varepsilon_{\rm e} = 0.01$, for given high-harmonic mode one appears two instability windows on the $k_z a$-axis, next to each other.  Thus, the range of KH instability is extended.  True, in the second instability windows the growth rates are much lower than in the standard instability windows, but nevertheless these weak/slow instabilities can occur.  Moreover, the critical flow velocity for emerging KH instability might be relatively lower.  For example, the $m = -4$ harmonic with wavelength of $11$~Mm can become unstable at jet speed of only ${\cong}25$~km\,s$^{-1}$ and its growth rate is equal to $0.001$~s$^{-1}$.  Solar H$\alpha$ surges are generally small-scale eruptive events and their contributions to solar coronal heating due to triggered by the KH instability wave turbulence is modest.  There are some cases, for instance as surges are detected in UV and EUV spectral lines, when the developing KH instability can bring forth a noticeable contribution to the solar corona energy budget.

\vspace{4.0mm}

\textbf{Acknowledgments} \hspace{0.5mm}
The works of I.Zh., R.C., A.K.S., and T.M.\ were supported by the Bulgarian Science Fund and the Department of Science \& Technology, Government of India Fund under Indo-Bulgarian bilateral project CSTC/INDIA 01/7, /Int/Bulgaria/P-2/12.  The work of T.Z.\ was supported by FP7-PEOPLE-2010-IRSES-269299 project- SOLSPANET, by Shota Rustaveli National Science Foundation grant DI/14/6-310/12 and by the Austrian Fonds zur F\"{o}rderung der wissenschaftlichen Forschung under project P26181-N27.  The authors are indebted to Dr.~Snezhana Yordanova for drawing one figure.

\end{document}